

\magnification=1200
\settabs 20 \columns

\+&&&&&&&&&&&&&&& UM-TH-92-32 \cr

\+&&&&&&&&&&&&&&& December 10,1992 \cr

\vskip2truecm


{\centerline {\bf Effective Lagrangian for $b \rightarrow s$ Processes }}
\medskip
{\centerline {\bf with QCD-Corrections }}


\vskip1.5truecm

\centerline { K.~ADEL and YORK-PENG YAO}
\bigskip

\centerline { \sl Randall Lab of Physics }
\medskip

\centerline { \sl University of Michigan }
\medskip

\centerline { \sl Ann Arbor, MI 48109 U.~S.~A. }

\vskip2truecm


\centerline{\bf Abstract }

\bigskip

for $b \rightarrow s$ processes with $QCD$ corrections,
in the leading logarithm approximation, for the cases
$m_t=m_w$ and $m_t \gg m_w$. The effective lagrangian is then applied
to estimate the $b \rightarrow s \gamma$ decay rate. We find that
it is enhanced by a factor of 6 for the case  $m_t=m_w$, and a factor
of about 2 for $m_t \gg m_w$. Our results differ from those
found in the literature, but are not very different
numerically for $m_t=m_w$.

\vfill \eject

\baselineskip=20pt
In the last three or four years a number of papers appeared$^{(1),(2)}$
which attempted
to include the heavy top quark effects in various flavor changing neutral
current (FCNC) processes.  The importance of these processes from a
theoretical point of view is their sensitivity to the top-quark mass
and other parameters such as Kobayashi-Maskawa Cabbibo (K-M) matrix
elements.  Indirect measurements of these parameters can be made by
studying low-energy processes such as $B-\bar B$ mixing and other rare
B-meson weak decays.  It is believed that rare decays, especially the
inclusive ones, will give stringent tests to the Standard Model (SM) since the
short distance quantum chromodynamics (QCD) effects can now be calculated
reliably in perturbation theory, independent of the long range effects, which
have a non-perturbative nature.
\smallskip
The short distance QCD effects are estimated in the framework of effective
theories, improved by renormalization group (RG) analyses.  Previous
calculations were done under the
assumption $m_w >> m_t$ or $m_t \sim m_w$, where $m_w$ and $m_t$ are
respectively the mass of the W-boson and that of the top-quark.  However,
recent analyses of the low energy phenomena indicate that the top quark
is in fact quite heavy.  Also, the CDF collaboration at Fermilab, which is
trying to detect the top-quark directly, has put a lower limit on the mass at
91 Gev ($90 \% $ confidence level.)  Besides, there are other
approximations made there, such as truncation of the "anomalous dimension"
matrix, and the use of the equations of motion$^{(1)}$, which are not in our
opinion
justified.  Furthermore, there have been a few calculations of the QCD-
corrections for the decay $b \rightarrow s \gamma \ (B \rightarrow K^\star
\gamma)$ which disagree with each other.  Except for disagreement in some
entries, our evaluation of the mixing matrix is quite close in spirit
to that by Misiak$^{(2)}$.
\smallskip
We have engaged in a program to account more properly for the top effects
enhanced by QCD, particularly to remove the assumptions and the approximations
mentioned earlier.  In this brief report, we shall use $b\rightarrow s
\gamma $ as an example to show how we perform our analysis.  It is important
to calculate correctly and reliably this decay, which is being measured at
CLEO (by J. Alexander et al.)  All this is
done in the context of effective theory, as formulated via Zimmermann
oversubstraction algebraic identities$^{(3)}$.
\smallskip
As is well-known, in many theories the heavy particle effects to the zeroth
order can be absorbed by reparameterization of the low energy Lagrangian.
This is the decoupling theorem.  However, the decoupling of heavy particles
does not always occur; weak interactions manifested with Yukawa couplings
are typical exceptions.  A very heavy top quark falls into this category.
(Worse yet, the removal of a member of a family at first sight renders the
effective theory unrenormalizable.)
Fortunately, even without decoupling it is still
possible to extract the heavy particle effects by including higher
dimension operators in the effective theory; the technique is the same as
that used to prove the decoupling theorem.  A set of renormalization group
equations (RGE) can be derived to effect a leading logarithmic sum of the
heavy particle effects for the Wilson coefficients,
with a few apparent assumptions.  We justify these
assumptions {\it aposteriori} by performing an
explicit exact two loop calculation of the Wilson coefficients, and then
compare with the results of the RGE.  We find that the large logarithms are
reproduced correctly.
\smallskip
In fact, one
cannot apply meaningfully a $RG$ analysis for the case $m_t \sim m_w$,
as was done previously.
The equation $m_t \sim m_w$ "suggests" that $m_t$ is
comparable to $m_w$,
but is not a very precise statement and leads to the
following problem concerning the solutions for the Wilson
coefficients: should we evaluate the initial conditions at $\mu=m_w$,
at $\mu=m_t$ or at some intermediate scale? We would like to emphasize that
the Wilson coefficients are very sensitive to these different choices of
initial conditions. Our answer to this {\it artificial} problem is simple:
if we do not assume anything between $m_t$ and $m_w$ more than just
that they are
large compared to the external momenta and light masses, then one has no
precise way to solve the $RGE$ for the Wilson coefficients
to reproduce the correct
results obtained by direct calculation in the full theory ($SM$).
This can be seen if one tries to compute explicitly two-loop Feynman diagrams
in the limit $m_t \sim m_w$: there, one obtains Spence functions whose
arguments are complicated square roots  of $m_t$, $m_w$ and light masses.
Expanding these Spence functions in a power series in $m_{light}/m_w$ and
$m_{light}/m_t$ does not $\underline {\rm always}$ generate simple
logarithms such as $ln(m_t/\mu)$. On the other hand the $RG$ analysis
in the $LLA$, when there is only one large scale, will always generate simple
logarithms!.  Thus, we shall consider only
either $m_t \gg m_w$ or $m_t$ equals exactly $m_w$.
For intermediate values of $m_t$ one can only make an {\it educated guess}.
\smallskip
As we shall see, even though there is some disagreement in the analytical
ouput, the difference numerically is not large for $b \rightarrow
s+\gamma $.  For the case $m_w=m_t$, we differ from Misiak's result by
about $1 \% $ and from Grinstein et. al. by $10\% $.  The rate is
enhanced by a factor of $\sim 6$.  The case for $m_t>>m_w$ is new and
the results may be used here for extrapolation as demanded or in 'new physics'.
\smallskip
Let us explain some more what we want to accomplish and how our approach is
different from the ones by others.  The full theory under discussion is the
SM, which is $SU(3)\times SU(2)\times U(1)$ invariant with three families
and one Higgs doublet.  The gauge-fixing term used here for the W-field is
($-C^+C^-$), where $C^+=-\partial _\mu W^+_\mu +m_w\phi^+ +ieA_\mu W^+_\mu .$
This choice reduces slightly the number of diagrams to be computed as well
as the number of operators needed for constructing the effective Lagrangian.
[It also makes the lowest oder effective Lagrangian $U(1)_{em}$ manifestly
invariant.]  The gauge-fixing term for the gluon is taken to be $-
(\partial _\mu G_\mu ^a)^2/2\alpha.$  The gauge parameter $\alpha $ is
arbitrary; it is there to check that the program of renormalization plus
others is carried out correctly.
\smallskip
In this report, we study the case $m_t \gg m_w$ in some details.  Our
final results will be valid for $m_t=m_w$ by choosing the appropriate
boundary conditions.
We want to integrate out {\it sequentially} the top quark field and then the
W-and $\phi $ boson fields.  As a first step we take the mass of the top
quark to be much heavier than any other scale in the theory.  The $W^{\pm }-$
and $\phi ^{\pm }-$bosons, the photon, the gluon, u,d,c,s and b quarks at
this point are treated as light fields.  Thus, in the limit $m_t>>m_w,
m_{light},p_{ext},$ we identify all heavy one-light-particle-irreducibe
(1LPI) Feynman diagrams contributing to order $m_t^0$ to $b \rightarrow s$
processes.  We discard all contributions $O(m_t^{-2})$ and smaller, such
as $m_w^2/m_t^2$; for $m_t\approx 2m_w$, the results for the Wilson
coefficients will have an error of roughly $25\% .$ It is in principle
possible to include these $m_w^2/m_t^2$ effects, but this requires
additional higher dimension (7 and 8) operators in the effective theory.
The intermediate effective Lagrangian is written as
$$L_{eff}^{(1)}=L_{light}^{(1)}+\sum _iC'_i(\mu )O_i^{(ren)}
+O({1\over m_t^2}), \eqno (1)$$
where $\lbrace O_i \rbrace$ is a set of local operators of dimension at
most equal to 6, made out of light fields only.  The prime is there to
remind us that the light theory and the operators still contain the heavy
W- and $\phi -$bosons.  The $\lbrace C'_i \rbrace $ are the Wilson
coefficients which contain all the heavy top quark effects.  Their
dependence is shown as $C'_i=C'_i(g_w,g_s,m_w,m_{light},m_t,ln(m_t^2/\mu ^2)).$
\smallskip
Let us briefly discuss how the $\lbrace O_i \rbrace $ are formed and the
$\lbrace C'_i \rbrace $ are extracted from a diagram.  Let $\Gamma $ be
a 1LPI diagram, which contains at least a top quark internal line.  Let
$\gamma $ represent a 1LPI graph or subgraph $(\gamma \subseteq
\Gamma )$ with external generic momentum p.  We define
$$\tau_\gamma^{(\epsilon )}=pole \ part \ of \ \gamma, \eqno (2)$$
$$\tau_\gamma^{(n)}=\gamma (p=0)+p{\partial \over \partial p}\gamma
(p=0)+ \cdot \cdot \cdot +{1\over n!}p^n{\partial ^n \over \partial
p^n }\gamma (p=0). \eqno (3)$$
The $R^\star $ renormalization scheme we use is defined as follows:
$$R^\star (\gamma _{heavy})=(1-\tau_\gamma ^{(n)})\gamma _{heavy}, \ \ \
  R^\star (\gamma _{light})=(1-\tau_\gamma ^{\epsilon})\gamma _{light}.
\eqno (4)$$
We then rearrange the $\gamma _{heavy} $ by applying the Zimmermann
oversubtraction identities.  These will be identified as $\Gamma (O_i)'s$
and the accompanying $C'_i.$  Power counting is then used to keep only
terms of $O(m_t^0)$.  The usefulness of the $R^\star $ scheme is that it
allows us to write the following equation between the Green's functions
in the full theory and those in the effective theory
$$\Gamma_{full}=\Gamma_{light}+\sum_iC'_i\Gamma_{light}(O_i), \eqno (5)$$
where the operators $O_i$ are inserted only once, and all the process
independent coefficients $C'_i$ are of order $m_t^0.$  This equation can then
be used to derive a RGE for the Wilson coefficients
$$\mu {dC'_i\over d\mu }=C'_j\gamma '_{ji} \ ; \ C'(\mu=m_t)=c'^0_i ,
\eqno (6)$$
where $\gamma'_{ji}$ is the anomalous dimension mixing matrix of the
operators $\lbrace O_i \rbrace$ and the $c'^0_i$ are the boundary
conditions. They can be calculated perturbatively.  Another feature of
the $R^\star $-scheme is that the renormalization group parameters are
those of QCD with five flavors.
\smallskip
Let us digress and see what would happen, had we used the minimal subtraction
scheme.  First, some of the coefficients associated with the dimension three
and four operators would be of order $m_t^2$.  This would not seem
troublesome, because they are also of order $g_w^2$ and therefore could not
be inserted more than once, as our analysis takes only lowest order weak
interaction into consideration.  However, the pure QCD sector would induce
operators such as $G^a_{\mu \nu}G^a_{\mu \nu}$ and $f^{abc}G^a_{\mu \nu}
G^b_{\mu \alpha }G^c_{\nu \alpha }$, which would enter into the analysis
because b and s mix under renormalization.   The first operator would have a
coefficient which is of order $m^0_t$ and could be inserted an infinite number
of times without being suppressed by a power of $m_t$.  The second operator
would have a Wilson coefficient of order $m_t^{-2}$ and could be inserted once
if it would be multiplied by a coefficient of order $m_t^2$.  For both
cases, it would be difficult to write a simple equation as Eq.(5) to derive a
RGE as Eq.(6)
for the Wilson coefficients.  The $R^\star -$scheme on the other hand makes
sure that no coefficients can be of order $m_t^2$ and that all the pure
QCD operators are multiplied by Wilson coefficients of order $m_t^{-2}$
and therefore can be discarded from this analysis.
\smallskip
The number of dimension five and six operators is very large.  This is
because gauge invariant operators can mix with non-gauge-invariant
operators which are allowed by the BRS symmetry.  The amount of effort
to extract $\gamma'_{ij}$ is quite substantial.  Nonetheless,
we follow the formulation of gauge theory operators by
Jogleckar and Lee, and others$^{(4)}$, for a complete operator basis to obtain
unambiguous decomposition.
\smallskip
The mixing between some operators is given by computing the infinites
of two-loop graphs.  There was some confusion about what prescription to
use for $\gamma ^5$ when calculating traces of Dirac matrices.  In our
work, there is no need to define a prescription for $\gamma ^5$ other
than that $\gamma $-algebra can be done in 4-dimensions in expressions
like $Tr(\gamma ^\mu \cdot \cdot \cdot )\gamma ^\alpha \cdot \cdot \cdot /
\epsilon $, at least for leading logarithm approximation (LLA).  We
find that the mixing of gauge invariant operators always involves
the calculation of simple pole parts of these quantities.
We use dimensional regularization and perform
all our calculations with a fully anti-commuting $\gamma ^5$ and justify
this procedure by comparing the results of the
RGE with an explicit two-loop calculation.  They are found
to be in complete agreement.
\smallskip
Among the dimension five and six operators, we single out a few for
this article.  These are (d= 5,6)
$$O_{61} =\, ({\bar{s}_{\alpha}}
{\gamma^{\mu}} {P_{\!\!{_L}}}
{c_\alpha} )
\,({\bar{c}_{\beta}}
{\gamma^{\mu}} {P_{\!\!{_L}}}
{b_\beta} ), \ \
O_{62} =\, ({\bar{s}_{\alpha}}
{\gamma^{\mu}} {P_{\!\!{_L}}}
{c_\beta} )
\,({\bar{c}_{\beta}}
{\gamma^{\mu}} {P_{\!\!{_L}}}
{b_\alpha} ),$$
$$O_{65} =\, ({\bar{s}_{\alpha}}
{\gamma^{\mu}} {P_{\!\!{_L}}}
{b_\alpha} )
\,({\bar{u}_{\beta}}
{\gamma^{\mu}} {P_{\!\!{_L}}}
{u_\beta}
+\ldots
+{\bar{b}_{\beta}}
{\gamma^{\mu}} {P_{\!\!{_L}}}
{b_\beta} ),$$
$$O_{63}=O_{61}(\bar c \rightarrow \bar u), \ \
O_{64}=O_{62}(\bar c \rightarrow \bar u), \ \ O_{66}=O_{65}((\alpha \alpha)
(\beta \beta )\rightarrow (\alpha \beta )(\beta \alpha )),$$
$$O_{67}=O_{65}(({P_{\!\!{_L}}} {P_{\!\!{_L}}})\rightarrow
({P_{\!\!{_L}}} {P_{\!\!{_R}}}), \ \
O_{68}=O_{66}(({P_{\!\!{_L}}} {P_{\!\!{_L}}})\rightarrow
({P_{\!\!{_L}}} {P_{\!\!{_R}}})),$$
$$ F_{51} = {\bar{s}}
{\phi^{+}}{\phi^{-}}
{({m_{s}\,P_{\!\!{_L}} + m_b\,P_{\!\!{_R}}})}b , \ \ \
 F_{62} = {\bar{s}_{_L}}{\lbrace}
{\phi^{+}}{\not\!\!D}{\phi^{-}}
{\rbrace}{b_{_L}}, $$
$$ F_{61} = {\bar{s}}{ \lbrace ({\not\!\!D+m_s})\,}
{\phi^{+}}{\phi^{-}}{P_{\!\!{_L}}}
+{P_{\!\!{_R}}}{\phi^{+}}{\phi^{-}}
{({\not\!\!D+m_b}) \rbrace}b ,$$
$$O_{51}=ig_s\bar sG_{\mu \nu}\sigma _{\mu \nu}(m_s{P_{\!\!{_L}}}+
m_b{P_{\!\!{_R}}})b/2,$$
$$O_{52}=ieQ_d\bar sG_{\mu \nu}\sigma _{\mu \nu}(m_s{P_{\!\!{_L}}}+
m_b{P_{\!\!{_R}}})b/2, \ \
\ Q_d=-1/3, \eqno (7)$$
and $W_{dj}$="2-quark" operators with $W^+W^-.$
When QCD-corrections are added, terms with large
"$g_s^2ln(m_t^2/\mu ^2)$" appear
which contribute to the $C'_i(\mu)$.  These make perturbation theory
unreliable.  We want to sum up all these large logarithms using the
RGE Eq.(6).  In LLA for
$m_t>>m_w$ and other scales, we can write
$$ C_i'(\mu) = \> g_w^2\> \sum_n \,a_n^{(i)}
\left( \> g_s^2(\mu)\> ln{m_t^2 \over \mu^2}\> \right)^n, \eqno (8) $$
and solve for the $a_n^{(i)}$.  We neglect terms which contain more
powers of $g_s^2$ than powers of $ln(m_t^2/\mu ^2)$.  This approximation
is justified if $\alpha _s$ is small compared to $\alpha _sln(m_t^2/\mu ^2)$.
Now, Eq.(6) tells us that if we compute $\gamma'_{ij}$ to order $g_s^2$, then
all the $a_n^{(i)}$ are related to the $a_0^{(i)}$, which are related to
the initial conditions $C'_i(m_t)=g_w^2a_o^{(i)}=c'^0_i$:
$$ c'^0_{O_{51}} \> = \> { G_t \over 16 \pi^2 } { 5 \over 12 }, \ \ \
c'^0_{O_{52}} \> = \> { G_t \over 16 \pi^2 } \left( -{11 \over 6 } \right)
  $$
$$ c'^0_{F_{61}} \> = \> - \> c'^0_{F_{62}} \> = \> -\  G_t, \ \ \
c'^0_{F_{51}} \> = \>  0 , \eqno (9)$$
$$ c'^0_{O_i} \> = \> 0, \  for \ O_i= O_{6j}, \ W_{dj},\ $$
where $G_t \, = \, { 2 \sqrt {2} } G_{\!F}  V_{ts}^*V_{tb},$ and
$G_{\!F}= \sqrt {2} \, g_w^2 /( 8\,m_w^2) $
is the Fermi coupling constant.  The parameter $m_t$ which appears in the
logarithm is the renormalized mass $m_t(\mu).$  However, in the $R^\star -$
scheme, in LLA, $m_t(\mu )=$constant.  We therefore choose $m_t=
(m_t)_{physical}.$
\smallskip
We first solve for the $\{C_i'\}$ corresponding to the "4-quark" operators
$O_{6i}$. These coefficients form a closed system, {\it i.e.}
they do not mix with the $C'_i$ associated with the "2-quark" operators.
This is because
$$ \gamma'_{ij}=0 \>\> if  \cases{i\neq O_{6k},& and \cr
                                             j=O_{6l},\> \cr} \eqno (10)$$
Regardless of how these 4-quark operators mix with themselves, they vanish
identically because their boundary conditions vanish. The same arguments
apply for the $\{C_i'\}$ associated with the operators $W_{dj}$, and we obtain
$$ C'_{O_i}(\mu)=0 \>\> \ if \ O_i=O_{6j},\>W_{dj}. \eqno (11)$$
We now focus on the $\{C_i'\}$ associated with the
operators $F_{dj}$,  $O_{51}$ and $O_{52}$. The mixing of these
operators with the "4-quark" operators is irrelevant at this stage
because of Eq.(11).  The mixing between themselves are found.
These coefficients form a closed system, and
as solutions their values at the scale $\mu$ are listed below:
$$ C'_{F_{di}}(\mu) \> = \> C'_{F_{di}} (m_t) \> , \ \ \
C'_{O_{51}}(\mu)\> = \> C'_{O_{51}} (m_t) \> \eta^{-{14 \over 3 \beta_
{\!F}} },$$
$$ C'_{O_{52}}(\mu)\> = \> \bigl(\> C'_{O_{52}} (m_t)
  + 8 \> C'_{O_{51}} (m_t) \> \bigr)
  \> \eta^{-{16 \over 3 \beta_{\!F}} }
  \,-\, 8 \> C'_{O_{51}} (m_t) \> \eta^{-{14 \over 3 \beta_{\!F}} }.
\eqno (12)$$
The following notation has been used:
$\beta_{\!F} \,=\, 11 - {2 \over 3 } N_f$,
where $ N_f=\,5$ is the number of flavors in the light theory,
$g_s(\mu)$ is the running coupling in $QCD$, and
$$ \eta \> = \> g^2_s(\mu)  / g^2_s(m_t)  \>
   = \> 1 + {\beta_F \over 16\pi^2}\, g_s^2(\mu)\, ln{m_t^2 \over \mu^2}.
\eqno (13)$$
\smallskip
Eqs. (11) and (12) together with the initial conditions Eq.(9) are our
results for the intermediate effective theory. It still contains the W- and
$\phi $-fields and will generate "heavy" graphs when calculating Green's
functions of light fields.  The heavy graphs are now the ones having at least
one internal W- or $\phi $-line, and "light" now refers to u, d, s, c, b,
$\gamma $ and gluon fields.  We want to integrate out the W- and $\phi $
-fields and obtain an effective theory which is made out of light fields only:
$$L_{_{eff}} = L_{_{light}}
     +\sum_i \,\,C_i(\mu)\, O_i^{(ren)}
     + O \big( {1 \over m_w^4} \big), \eqno (14)$$
valid for $m_w>>\mu, p_{ext}, m_{light}.$   All the $1/m_w^2$ effects
are now contained in the new Wilson coefficients $C_i(\mu )$.  Here as
before our task is to perform a LLA for
$$C_i(\mu) = g_w^2 \, \sum_{n,m=0}^{\infty}
 A_{nm}^{(i)}
\left[ g_s^2\,ln{m_t^2 \over \mu^2} \right]^n \,
\left[ g_s^2\,ln{m_w^2 \over \mu^2} \right]^m. \eqno (15) $$
The $C_i$ satisfy the same RGE as Eq.(6), with the primes removed.  However,
the operators $W_{dk}$ and $F_{dk}$ have disappeared.  The mixing
$\gamma _{ij}$ is the same as $\gamma '_{ij}$ for $i,j=O_{6k}, O_{51},$
and $O_{52}$.  The boundary conditions $C_i(\mu =m_w)=c_i^0$ are
$$C_{O_{51}}(m_w) \> = \>  { G_t \over 16 \pi^2 }
\left[
\>  -{2 \over 3} \, +\, { 5 \over 12 } \, \eta_0\,^{- {14 \over 3 \beta_{F}}}
\right] ,$$
$$C_{O_{52}}(m_w) \> = \>  { G_t \over 16 \pi^2 }
\left[
\>  {23 \over 6} \, + \,{ 3 \over 2 } \, \eta_0\,^{- {16 \over 3 \beta_{F}}}
\, -\, { 10 \over 3 } \, \eta_0\,^{ - { 14 \over 3 \beta_{F} } }
\right] ,  \eqno (16)$$
$$C_{O_{61}} (m_w) \> = \> -\,  G_c , \ \ \
C_{O_{63}} (m_w) \> = \> -\,  G_u ,$$
$$C_{O_{6k}} (m_w) \> = \> 0 \ \ otherwise $$
where
$  G_c \, = \, { 2 \sqrt {2} } G_{\!F}  V_{cs}^*V_{cb} , \ \
 G_u \, = \, { 2 \sqrt {2} } G_{\!F}  V_{us}^*V_{ub} $
and $ \eta_0 =\eta (\mu=m_w)$.
The up (u) and the charm (c) quarks now enter into the boundary
conditions.  We have all the necessary mixing to solve the RGE in the LLA
with no further approximation.  In particular, we do not need to truncate
the mixing matrix as was done by previous authors.  Because of the structure
of the mixing matrix, we can write the RGE into four sets:
$$ \eqalign{
\mu {d \over{d\mu}}\,C_{O_{61}} \> &= \>
{ g^2_s \over 8 \, \pi^2 } \>
\left(\>
-\,C_{O_{61}} \,+\, 3 \,C_{O_{62}} \>
\right), \cr
\mu {d \over{d\mu}}\,C_{O_{62}} \> &= \>
{ g^2_s \over 8 \, \pi^2 } \>
\left(
\> 3\,C_{O_{61}} \,-\, C_{O_{62}} \>
\right),\cr } \eqno(17)$$
$$\mu {d \over d\mu} \,\vec{C}^t \>  = \>
{ g^2_s \over 8 \, \pi^2 } \> \left[ \>
\vec{C}^t \,T \, + \, \vec{A_0}^t \,
\left(\, C_{O_{61}} \,+\, C_{O_{63}} \, \right) \, \right], \eqno (18)$$
where $T$ is given below for  $N_f=5$
$$T\>= \>
\pmatrix{
 -11/9 & 11/3 & -2/9 &  2/3  \cr
  22/9 &  2/3 & -5/9 &  5/3  \cr
    0  &   0  &   1  &   -3  \cr
  -5/9 &  5/3 & -5/9 & -19/3 \cr
       },\eqno (19)$$
$$ \eqalign{
\vec{C}^t \,&=\,\left(
\> C_{O_{65}} \;,\; C_{O_{66}} \;,\; C_{O_{67}} \;,\; C_{O_{68}} \>
\right), \cr
\vec{A_0}^t \,&=\,
\left( \> -1/9 \;,\; 1/3 \;,\; -1/9 \;,\; 1/3 \> \right),\cr}$$
$$\eqalign{
\mu {d \over{d\mu}}\,C_{O_{51}} \> = \>
{ g^2_s \over 8 \, \pi^2 } \> \Bigl[\>
&\,+\, {14 \over 3 }\,C_{O_{51}} \,
+\,{19 \over 108 \pi^2 }\,\left(\,C_{O_{61}}\,+\,C_{O_{63}}\,\right) \cr
&+\,{3 \over 16 \pi^2 }\,\left(\,C_{O_{62}}\,+\,C_{O_{64}}\,\right) \,
+\,{1\over \pi^2}\,\vec{A_1}^t \vec{C} \> \Bigr], \cr } \eqno (20)$$
with
$$\vec{A_1}^t \>=\>\left( \> {557\over 432} \>,\>
{271\over 216} \>,\> -\,{53\over 48} \>,\>
-\,{457\over 432} \> \right),  $$
and
$$ \eqalign{
\mu {d \over{d\mu}}\,C_{O_{52}} \> = \>
{ g^2_s \over 8 \, \pi^2 } \>
\Bigl[\>
\,&+\,{16 \over 3 }\,C_{O_{51}} \, +\,{16 \over 3 }\,C_{O_{52}} \cr
&-\,{29 \over 27 \pi^2 }\,\left(\,C_{O_{61}} \,+\,C_{O_{63}} \,\right)\,
+\,{1\over \pi^2} \,\vec{A_2}^t \vec{C} \> \Bigr], \cr } \eqno (21)$$
with
$$ \vec{A_2}^t \>=\>\left( \> {23\over 27} \>,\>
-\,{47\over 54} \>,\> {13\over 3} \>,\>
{493\over 54} \> \right). $$
\smallskip
We solve Eqs.(17), (18), (20), and (21) successively with the boundary
conditions of Eq.(16).  This involves the diagonalization of the matrix
T of Eq.(19), which is done numerically.  After applying the unitarity
relation $G_t+G_c+G_u=0$, the result we are after is
given as:
$$ \eqalign{
C_{O_{52}}(\mu) =\;\; \eta_w^{-16/3\beta_{_F}} \> \big[\,
& C_{O_{52}}(m_w) \,+\,8\,C_{O_{51}}(m_w) \,
+\, (8.009)\,{G_t\over 16\,\pi^2} \,\big] \cr
+\,\eta_w^{-14/3\beta_{_F}} \> \big[\,
& -\,8\,C_{O_{51}}(m_w) \,
-\, (1.952)\,{G_t\over 2\,\pi^2} \,\big] \cr
+\,\eta_w^{-2/\beta_{_F}} \> \big[\,& {18\over 7}\,
{G_t\over 16\,\pi^2} \,\big]
+\,\eta_w^{4/\beta_{_F}} \> \big[\, {3\over 7}\,
{G_t\over 16\,\pi^2} \,\big] \cr
+\,{G_t\over 16\,\pi^2} \, \big[\>&
+\,(0.63)\,\eta_w^{-a_1/\beta_{_F}} \,
-\,(0.165)\,\eta_w^{-a_2/\beta_{_F}} \cr
& -\,(0.187)\,\eta_w^{-a_3/\beta_{_F}} \,
+\,(4.327)\,\eta_w^{-a_4/\beta_{_F}} \>\big]\cr },\eqno(22)$$
with $a_1=-6.90, \ a_2=-3.24, \ a_3=1.12, \ a_4=3.13$ and
$\eta _w=\eta (m_t\rightarrow m_w)$ in Eq.(13).  As with Eq.(12),
we have explicitly verified the results with a two loop calculation.
We can perform an analysis for the case $m_w=m_t$ in the same fashion.
The only major change is the boundary conditions.
\smallskip
To give an estimate of the magnitude of the $QCD$
corrections to the Wilson coefficients, we look at Figure 1,
where we have displayed the $m_t$-dependence
of $\bar{C}_{O_{52}}(m_b)=16\pi^2 C_{O_{52}}/G_t$.
Let us briefly explain what the different curves correspond to:
\noindent
(i) The dot-dashed curve, which is labeled "$NO$ $QCD$
($m_t\sim m_w$)" represents the lowest order Wilson coefficient (also called
the Inami-Lim function$^{(5)}$), obtained
by assuming that $m_t \sim m_w \gg p_{ext},\,m_{light}$.
(ii) The black disk on the vertical axis at $m_t/m_w=1$, which is labeled
"$QCD$ ($m_t=m_w$)", represents
the value of the $QCD$-corrected Wilson coefficient at that point.
(iii) The curves enclosed in brackets represent the result for the Wilson
coefficient which was obtained in the limit $m_t \gg m_w$.
Note that we are assuming that the result obtained in this limit can
be used as a first approximation (within 25 to 30 \%) to the case
$m_t \geq 2\,m_w$. For this case, the solid curve which is labeled
"$QCD$" is our $QCD$-corrected result for $\bar{C}_{O_{52}}$;
the dashed curve, which is labeled "$QCD$-truncated",
represents the result which would be obtained if the anomalous mixing matrix
were truncated, as was done by Grinstein et al$^{(1)}$.
Finally the dotted line which is labeled "$NO$ $QCD$" is the
initial condition obtained for $m_t\gg m_w\ ^{\dagger}$.
The numerical work was done for $\alpha_s(m_b)=0.19$.

\smallskip
For $b\rightarrow s+\gamma$ we have to take into account the {\it on-shell}
effects coming from
the contribution of the four-quark operators$^{(2)}$:
this amounts to redefining
an {\it on-shell} effective Wilson coefficient $\bar {C}_{O_{52}}^{eff}
=16\pi ^2C_{O_{52}}^{eff}/G_t$, with
$$ C_{O_{52}}^{\,eff}(m_b) \,=\,
                         C_{O_{52}}(m_b)
  \,+\, {1\over 8 \pi^2} C_{O_{67}}(m_b)
  \,+\, {3\over 8 \pi^2} C_{O_{68}}(m_b)\eqno (23).$$
To remove the dependence on $|G_t|^2$, which is not accurately
known experimentally, we normalize the
$b \rightarrow s \,\gamma$ partial width to the well established
semileptonic $b \rightarrow ce\bar{\nu} \,$ partial width, and use the
following relation: $|V_{ts}^*V_{tb}| \simeq |V_{cb}|.$
This ratio is given as:
$$ {\Gamma \left(b \rightarrow s \gamma\right)
\over \Gamma \left( b \rightarrow c e \bar{\nu} \right) } \simeq
{ \alpha_{QED} \over 6 \pi \, g\,(m_c/m_b)}\,
\left( 1 - {2\alpha_s(m_b) \over 3 \pi} f(m_c/m_b) \right)^{-1}\,
|\bar {C}_{O_{52}}^{\,eff}(m_b)|^2 ,\eqno (24)$$
where $g(m_c/m_b)\simeq 0.45$ and $f(m_c/m_b)\simeq 2.4$ correspond to
the phase space factor and the one-loop $QCD$ corrections to the
semileptonic decay, respectively.
Without attempt to interpolate,
the ratio is $2.2\times 10^{-3}$ for $m_t=m_w$ and
about $7\times 10^{-3}$ for $m_t \gg m_w.$  The $QCD$-enhancement factor for
the decay $b\rightarrow s \gamma$ is found to be large:
6 for the case $m_t=m_w$, and about 2 for the case $m_t \gg m_w$.  We
are compelled to remind the reader also that the values of the relevant
Wilson coefficients without QCD for $m_t \gg m_w$ are substantially
larger than those for $m_t \sim m_w$.
\smallskip
Further details of this article will be published elsewhere. We have
extensively used the symbolic algebraic program Schoonschip$^{(6)}$ in
this work.
\noindent
This work has been partially supported by the U.~S.~ Department of Energy.

\vfill \eject
\noindent
{\it FOOTNOTES }

\bigskip
\+ $^{\dagger}$ & In view of what we have done, one way to
                  extrapolate is to use the \cr
\+              & combination $QCD(m_t \gg m_w) + \left(\,NO\;
                  QCD(m_t \sim m_w) - NO\;QCD(m_t \gg m_w) \,\right)$. \cr
\+              & This will honestly take into $QCD$ effects
                  for $O(m_t^0)$ and $O(m_w^{-2})$ terms and \cr
\+              & include $O(m_t^{-2})$ and higher effects without $QCD$.
                  By the same token,  \cr
\+              & $QCD(m_t=m_w) + \left(\, NO\; QCD(m_t \sim m_w)
                  - NO\;QCD(m_t=m_w) \,\right)$ will do as well.  \cr
\+              & These two approaches give for $m_t = 2 m_w$,
                  $\bar {C}_{O_{52}}(m_b)=1.75$ and $2.06$ respectively. \cr

\bigskip
\noindent
{\it REFERENCES }

\bigskip

\settabs 20 \columns

\+ [1] & B.~Grinstein, R.~ Springer and M.~B.~Wise, {\sl Phys. Lett.}
    {\bf 202 B} (1988) 138. \cr
\+     &  See also the references listed in these articles;  R.~Griganis,
P.~J.~O'Donnell,\cr
\+     & M.~Sutherland and H.~Navalet, {\sl Phys. Lett.} {\bf 213 B} (1988)
          355 ; N.~G.~Deshpande \cr
\+     & and J.~Trampetic, {\sl Mod. Phys. Lett.}
	{\bf A 4} (1989) 2095. \cr
\medskip

\+ [2] & M.~Misiak, {\sl Phys. Lett.} {\bf 269 B} (1991) 161;
M.~Misiak, {\sl preprint} ZH-TH-19/22 (1992). \cr

\medskip

\+ [3] & W.~Zimmermann, in {\sl Lectures in Elementary Particles and Quantum
        Field} \cr
\+ &  {\sl Theory}, edited by S.~Deser {\sl et al.}
    (MIT Press, Cambridge, Mass.,
      1971),\cr
\+ & Vol. I, p.397;
Y.~Kazama and  Y.-P.~Yao, {\sl Phys. Rev.} {\bf D 25} (1982) 1605.\cr

\medskip

\+ [4] &  S.~D.~Jogklekar and B.~W.~Lee, {\sl Ann. Phys.}
            {\bf 97} (1976) 160. \cr

\+     & See also, W.~S.~Deans and J.~D.~Dixon, {\sl Phys. Rev.}
            {\bf D 18} (1978) 1113, \cr

\+     & and  H.~Kluberg-Stern and J.~B.~Zuber, {\sl Phys. Rev.}
            {\bf D 12} (1975) 467. \cr

\medskip

\+ [5] & T.~Inami and C.~S.~Lim, {\sl Progr. Theor. Phys. } {\bf 65}
(1981) 297-314 \cr

\medskip

\+ [6] & Schoonschip program by M.~Veltman, unpublished. \cr

\vfill \eject

\noindent
{\it FIGURE CAPTION:}
\bigskip
\noindent
Figure 1: $C_{O_{52}}(m_b)$ dependence on $m_t$ with and without QCD
corrections.  See text for explanation of various curves.
\vfill \eject

\input epsf
\epsfxsize=\hsize

\vfill
\epsffile{/usr/kassa/tex/CV/figure6_1.eps}
\vfill

\centerline{ Figure 1.}
\vfill \eject
\end